\documentclass[twocolumn,superscriptaddress,prb,preprintnumbers,nobibnotes,aps]{revtex4-1}  
\usepackage{graphicx}
\usepackage[caption=false,font=large]{subfig}
\usepackage{floatrow}
\usepackage{epstopdf}
\usepackage{dcolumn}
\usepackage{hyperref}
\hypersetup{
  colorlinks  = true, 
  urlcolor   = blue, 
  linkcolor  = blue, 
  citecolor  = blue 
}
\usepackage{color}
\usepackage{soul}
\usepackage{natbib}
\usepackage{amsmath}

\newcommand{\CGT}{Cr$_2$Ge$_2$Te$_6$}

\begin{document}

\title{Magnetic Anisotropy and Low Field Magnetic Phase Diagram of Quasi\,Two-Dimensional Ferromagnet Cr$_2$Ge$_2$Te$_6$}

\author{S.~Selter}
\email{s.selter@ifw-dresden.de}
\affiliation{Institute for Solid State Research, Leibniz IFW Dresden, Helmholtzstr. 20, 01069 Dresden, Germany}
\affiliation{Institute of Solid State and Materials Physics, Technische Universit\"at Dresden, 01062 Dresden, Germany}
\author{G.~Bastien}
\affiliation{Institute for Solid State Research, Leibniz IFW Dresden, Helmholtzstr. 20, 01069 Dresden, Germany}
\author{A.~U.~B.~Wolter}
\affiliation{Institute for Solid State Research, Leibniz IFW Dresden, Helmholtzstr. 20, 01069 Dresden, Germany}
\author{S.~Aswartham}
\email{s.aswartham@ifw-dresden.de}
\affiliation{Institute for Solid State Research, Leibniz IFW Dresden, Helmholtzstr. 20, 01069 Dresden, Germany}
\author{B.~B\"{u}chner}
\affiliation{Institute for Solid State Research, Leibniz IFW Dresden, Helmholtzstr. 20, 01069 Dresden, Germany}
\affiliation{Institute of Solid State and Materials Physics, Technische Universit\"at Dresden, 01062 Dresden, Germany}
\date{\today}

\begin{abstract}
In this work we present a comprehensive investigation on magnetic and thermodynamic properties of the two-dimensional layered honeycomb system \CGT. Using magnetization and specific heat measurements under magnetic field applied along two crystallographic directions we obtain the magnetic phase diagram for both directions. \CGT\ is a ferromagnet with a Curie temperature $T_C=65$\,K and exhibits an easy magnetization axis perpendicular to the structural layers in the \textit{ab}-plane. Under magnetic fields applied parallel to the hard plane \textit{ab} below the magnetic saturation, a downturn with an onset temperature \textit{T*} is observed in the temperature dependent magnetization curve. \textit{T*} shows a monotonous shift towards lower temperatures with increasing field. The nature of this anisotropic and specific behavior for fields in the hard plane is discussed as an interplay among field, temperature and effective magnetic anisotropy. Similarities to structurally related compounds such as CrX$_3$ (X = Br, I) hint towards a universality of this behavior in ferromagnetic quasi two-dimensional honeycomb materials.
\end{abstract}

\maketitle

\section{Introduction}
Since the discovery of Graphene in 2004\cite{KSNovoselov2004}, two dimensional (2D) materials have been in the forefront of research both in fundamental as well as in applied science. This class of materials stands out due to novel electronic properties in combination with unique structural characteristics \cite{KFMak2010,ASplendiani2010,XXi2015,SKolekar2017,AWTsen2015}. On one hand, when thinned down to the monolayer limit, significant changes in the physical properties have been observed\cite{KFMak2010,ASplendiani2010,AWTsen2015,ASPawlik2018}. On the other hand, some materials conserve their bulk properties down to the monolayer limit, enabling new applications and architectures\cite{LDAlegria2014,KSNovoselov2016,AGeim2013}. Examples are ferromagnetic monolayers, which have a great potential for applications in the field of spintronics and data storage devices.\\
As observed in \CGT \cite{CGong2017} and in structurally related CrI$_3$\cite{BHuang2017}, evidence for ferromagnetism at least down to the bilayer could be seen by magneto-optical-Kerr-effect (MOKE) microscopy. The structural relation between \CGT\ and CrI$_3$ is given by a shared honeycomb motif in the \textit{ab}-plane. For the iso-structural compound Cr$_2$Si$_2$Te$_6$ monolayer ferromagnetism is theoretically predicted\cite{MLin2016} but still lacks experimental confirmation. The presence of magnetic anisotropy plays a crucial role in monolayer magnetism. As predicted in the Mermin-Wagner theorem\citep{NDMermin1966}, isotropic Heisenberg interactions in dimensions $\leq 2$ will be disturbed by long range fluctuations. However, taking into account already a weak anisotropy, the proof in the theorem is no longer valid and long range magnetic order may be stabilized in low dimensions. Furthermore, Kitaev interactions were recently discussed to realize the magnetic exchange mechanism in the monolayer of these compounds\citep{CXu2018}. It is also worth mentioning, that VSe$_2$, a diamagnet in bulk, shows ferromagnetic ordering when prepared as a monolayer\cite{MBonilla2018}.\\
While the discovery of robust ferromagnetism in the monolayer limit itself is without doubt stunning and attracted significant attention in the scientific community due to the potential impact it can have in future applications, the bulk magnetic state in these compounds is not well understood. For example, for all mentioned bulk ferromagnets, an anisotropic magnetic anomaly can be observed applying relatively low fields\cite{XZhang2016,MMcguire2015,YLiu2017}. Until now the origin and nature of this anomaly remains elusive. However, to entangle the physics behind the intriguing phenomenon of monolayer ferromagnetism, a reliable understanding of the bulk magnetism and anisotropy is a prerequisite in these compounds.\\

\begin{figure*}[!htbp]
\centering
\includegraphics[width=\linewidth]{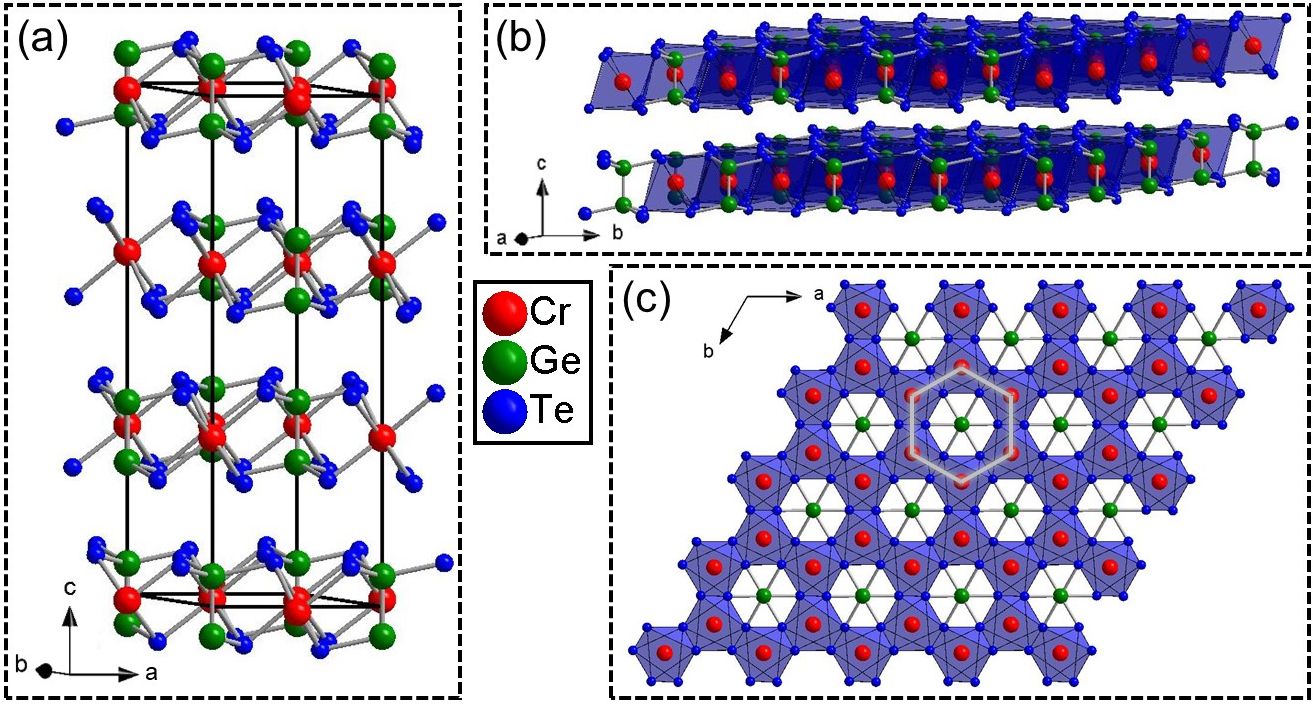}
\caption{Crystal structure of \CGT\ in the space group \textit{R}\={3} (No. 148). (a) Unit cell
of \CGT. (b) View perpendicular to the \textit{c}-axis showing the structural layers and
their stacking. (c) View perpendicular to the \textit{ab}-plane showing the honeycomb
network.} \label{fig:Crystal_Structure}
\end{figure*}

\CGT\ crystallizes in the trigonal space group \textit{R}\={3} (No.\,148) and belongs to the class of layered van-der-Waals (vdW) transition metal trichalcogenides (TMTC). This class of compounds possesses layers made of the respective transition metal (TM), octahedrally surrounded by the respective chalcogenide (C)\cite{VCarteaux1995,RBrec1986}. Those edge-sharing TMC$_6$ octahedra form a honeycomb network. The void of each honeycomb is occupied by a dimer of a IV/V main group element (P, Si, Ge) with the binding axis between the two atoms perpendicular to the honeycomb plane. This dimer is a peculiarity which differentiates this structure from other honeycomb structures, such as CrX$_3$ (X = Cl, Br, I). As shown in Fig.~\ref{fig:Crystal_Structure}(b), the honeycomb layers are stacked onto each other, well separated by a van der Waals (vdW) gap, which makes it easy to exfoliate crystals down to a few layers. The stacking of the layers varies in the family of TMTCs. For \CGT\ and Cr$_2$Si$_2$Te$_6$ in the \textit{R}\={3} space group (No.\,148), an ABC stacking is found. In contrast, Al$_2$Si$_2$Te$_6$ in the \textit{P}\={3} (No.\,147) space group (with a main group metal instead of a transition metal) exhibits the highly ordered AAA stacking\citep{ESandre1994}. For the TM$_2$P$_2$(S,Se)$_6$ family of compounds, the stacking is more difficult to generalize, since the stacking of the layers with respect to a perpendicular direction depends on the monoclinic $\beta$ angle of the space group \textit{C}12/\textit{m}1 (No.\,12) \citep{AWildes2015,GOuvrard1985}. These considerations of the stacking do not explicitly take stacking faults into account.\\

TMTCs in general possess a non-zero bandgap ranging from 0.5\,eV to 3.5\,eV mainly depending on the TM and the strong spin-orbit coupling together with electron correlations\cite{AMishra2018}. Furthermore, these compounds exhibit many different possibilities for long-range magnetic order, mainly depending on the TM ion. \CGT\ in particular has a bandgap of $\sim$\,0.74\,eV (direct) and $\sim$\,0.2\,eV (indirect) and a ferromagnetic ground state with the magnetic easy axis perpendicular to the layers\cite{VCarteaux1995,HJi2013}. This makes the title compound one of the rare examples of ferromagnetic semiconductors. Owing these properties and the nature of this class of materials to be easy to exfoliate, \CGT\ found use as substrate for ferromagnetic insulator-topological insulator heterostructures\cite{LDAlegria2014}. Furthermore, the magnetic lattice of \CGT\ (and also Cr$_2$Si$_2$Te$_6$) is the same as for CrX$_3$ (X = Br, I), since the Ge dimer in the void of the \CGT\ honeycomb is magnetically inactive. Altogether, the known 2D vdW honeycomb ferromagnets exhibit an excellent platform to compare their magnetic interactions.\\

Here, we present a comprehensive experimental investigation of the anisotropic bulk magnetic properties of vdW-layered \CGT\ single crystals by means of DC magnetometry and specific heat measurements. We obtain the low-field magnetic phase diagram of this compound for the easy axis and
hard plane, with the easy-axis being perpendicular to the honeycomb layers. Under magnetic fields applied parallel to the hard plane \textit{ab}, a downturn with an onset temperature \textit{T*} is observed in the temperature dependent magnetization curve. We explain this anisotropic and specific behavior for fields in the hard plane as an interplay among field, temperature and effective magnetic anisotropy in \CGT.

\section{Synthesis, Sample Characterization and Methods}
\label{Samples_and_Methods}

\begin{figure}[!htbp]
\centering
\includegraphics[width=0.5\linewidth]{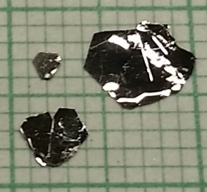}
\caption{As-grown crystals of \CGT\ up to several mm size.} \label{fig:Image_CGT_crystals}
\end{figure}

Single crystals of \CGT\ with a size up to 6\,mm\,x\,5\,mm\,x\,0.2\,mm (see Fig.~\ref{fig:Image_CGT_crystals}) were grown by the self flux technique according to X. Zhang \textit{et al.} \citep{XZhang2016}. Details regarding the growth procedure and an in-depth characterization of the crystals used in this work are published elsewhere \citep{JZeisner2019}. Both powder X-ray diffraction and energy dispersive X-ray spectroscopy agree well with the published crystal structure in the R\={3} space group\citep{VCarteaux1995} as well as with the expected stochiometry of \CGT.\\

DC magnetization was measured as a function of temperature and field using a quantum interference device vibrating sample magnetometer (SQUID-VSM) from Quantum Design. The values obtained for magnetic moments were corrected due to deviation of the measured sample shape and size from a point dipole. This correction follows the procedure described in Ref.~\cite{VSM_corr}. A detailed description of how this correction is applied can be found in the Appendix of the work of J. Zeisner \textit{et al.}\citep{JZeisner2019}.\\
Low-temperature specific heat was determined using a relaxation technique in a Physical Property Measurement System (PPMS) from Quantum Design. The specific heat from the platform and grease used for mounting the sample were subtracted.

\section{Results and discussion}
\label{Results_Discussion}

\subsection{Magnetic Characterization}

\begin{figure}[!htbp]
\centering
\includegraphics[width=\linewidth]{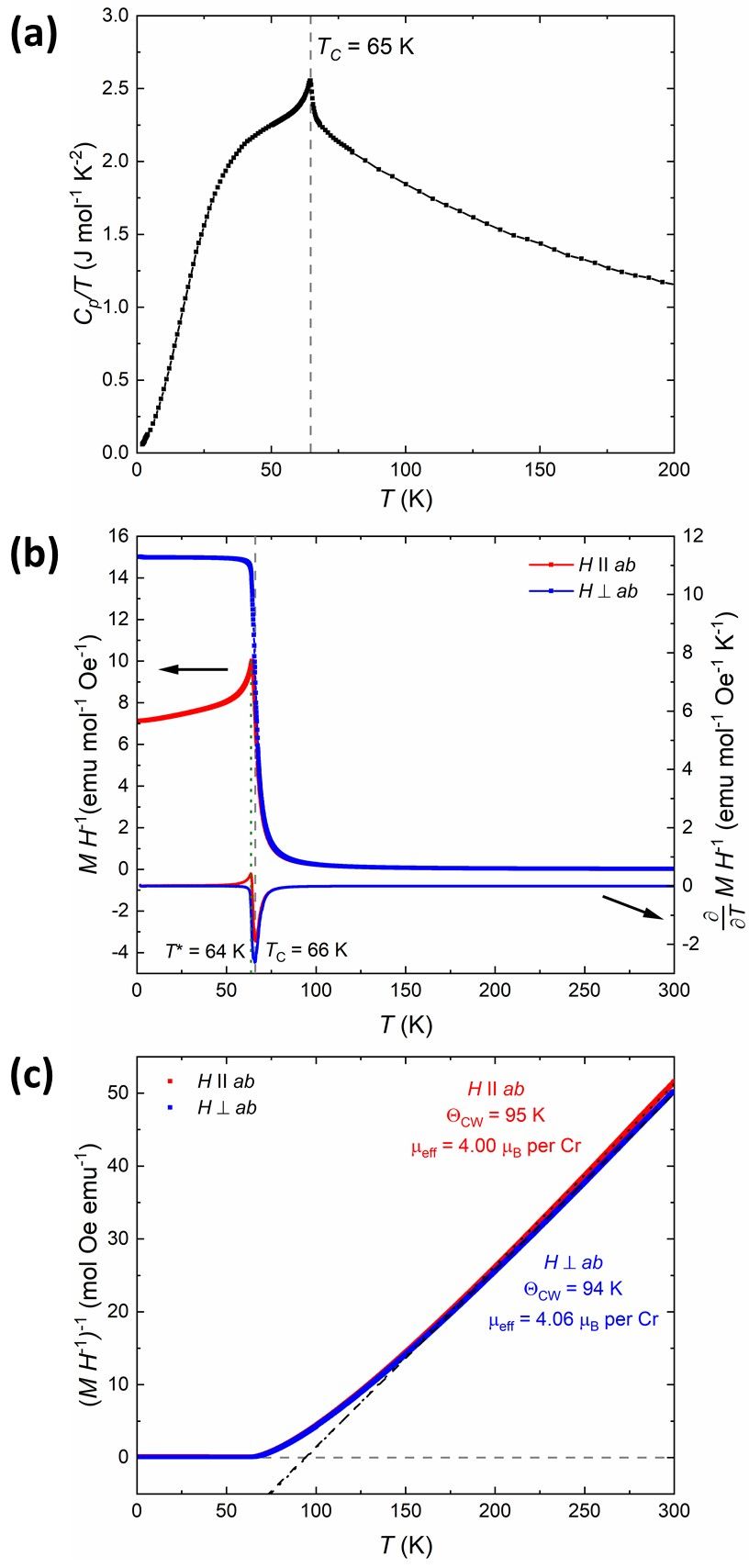}
\caption{(a) Zero-field specific heat divided by the temperature $C_p/T$ of Cr$_2$Ge$_2$Te$_6$ as a
function of temperature. (b) Temperature dependence of the normalized magnetization \textit{M/H}
and its first derivative at \textit{H}\,=\,1\,kOe (field-cooled). The grey dashed line indicates
the Curie temperature $T_C$ and the green dotted line indicates the temperature of the onset of the
downturn \textit{T*} for $H \parallel ab$. (c) Inverse of the normalized magnetization
($(M/H)^{-1}$) at \textit{H}\,=\,1\,kOe. The black dashed and dotted lines show linear fits in the
paramagnetic region (150~K $<$ $T$ $<$ 300~K).} \label{fig:basic_magnetic_characterization}
\end{figure}

The zero-field specific heat of Cr$_2$Ge$_2$Te$_6$ divided by temperature $C_p/T$, and the
temperature dependent normalized magnetization \textit{M/H} at 1\,kOe applied parallel and
perpendicular to the crystallographic \textit{ab}-plane as well as the inverse of the normalized
magnetization are represented in Fig.~\ref{fig:basic_magnetic_characterization}(a), Fig.~\ref{fig:basic_magnetic_characterization}(b) and
Fig.\,\ref{fig:basic_magnetic_characterization}(c), respectively. For the normalized magnetization only the results
from field-cooled measurements are shown since no significant difference of zero-field-cooled and
field-cooled measurements was observed.\\

A $\Lambda$-shape peak in the temperature dependent
specific heat indicates a second-order phase transition at $T_C=65$\,K. In good agreement with
this, a similar Curie temperature ($T_C = 66\,(\pm\,1)$\,K) is obtained from the minimum of the
first derivative of the temperature dependent normalized magnetization for both crystallographic
orientations. While no further phase transition was observed in the specific heat at zero field, the
magnetization curves shows an anomalous behavior for $H \parallel ab$ below $T_C$. A downturn
towards lower \textit{T} is observed below \textit{T*}\,=\,64\,K for $H \parallel ab$, whereas for
$H \perp ab$ a typical ferromagnetic behavior is observed.\\
A similar anisotropic behavior is also
seen for Cr$_2$Si$_2$Te$_6$\citep{LCasto2015}, CrI$_3$\citep{NRichter2018} and
CrBr$_3$\citep{NRichter2018}, which are also 2D honeycomb ferromagnets and which show a close
relation to \CGT\ regarding their structure. The similarities regarding structure, magnetic ion and
magnetic ordering hint towards a main role of these properties for the origin of the observed
anisotropy.\\

At temperatures well above the Curie temperature in the paramagnetic state, a linear dependence
between magnetization and field can be assumed. Therefore the magnetic susceptibility can be
approximated by the normalized magnetization as shown in Eq.~\ref{eq:chi_MH_relation}.
\begin{equation}
\chi(T) = \frac{\partial M}{\partial H} \approx \frac{M}{H}.
\label{eq:chi_MH_relation}
\end{equation}
Consequently, in the paramagnetic state the normalized magnetization can be used for a Curie-Weiss
analysis. From this analysis effective magnetic moments of $\mu_{eff}$\,=\,4.00\,$\mu_{B}$/Cr for
$H \parallel ab$ and $\mu_{eff}$\,=\,4.06\,$\mu_{B}$/Cr for $H \perp ab$ are obtained, which is in
good agreement with the theoretically expected spin-only moment of $\mu_{so} = 3.87 \mu_B$ for Cr$^{3+}$. Furthermore,
our Curie-Weiss analysis yields a Curie-Weiss temperature of $\Theta_{CW}$\,=\,95\,K for $H
\parallel ab$ and $\Theta_{CW}$\,=\,94\,K for $H \perp ab$ in good agreement with literature\citep{YLiu2017,GLin2017}.\\
The positive Curie temperature indicates a dominant ferromagnetic coupling. In three-dimensional
ferromagnets $\Theta_{CW}$ is generally found to be close to $T_C$. The difference between
$\Theta_{CW}$ and $T_C$ that is found for \CGT\ is most likely an indication for the suppression
of the magnetic order due to the two-dimensional nature of the compound and thus also of the
magnetic interactions. This is in line with current results obtained from ferromagnetic resonance
(FMR) and electron spin resonance (ESR)\citep{JZeisner2019}, which demonstrated the intrinsic
two-dimensional nature of the magnetic interaction in \CGT. Also the temperature dependence of
$C_{p}/T$ in Fig.~\ref{fig:basic_magnetic_characterization}(a) shows characteristic features for the two-dimensional
nature of the magnetic interactions in \CGT: the $\Lambda$-shape peak is rather small with an estimated
integral of approximately $\Delta S_\Lambda \simeq 2~$J/mol/K compared to the expected value of the magnetic
entropy change at a ferromagnetic ordering of a system with two $S=3/2$ magnetic ions per unit
cell, the latter being $S_{mag}=2Rln(4)=23.05~$J/mol/K. This indicates, that the broad bump in the
experimentally determined $C_p/T$ contains a sizable magnetic contribution in addition to the
phononic contribution. Thus magnetic fluctuations give an important contribution to the specific
heat even far above and far below the magnetic ordering. This is certainly related to the quasi
two-dimensional nature of the magnetism in Cr$_2$Ge$_2$Te$_6$, as previously proposed by
G.\,T.\,Lin\,\textit{et\,al.}\cite{GLin2017}.

\begin{figure}[!htbp]
\centering
\includegraphics[width=\linewidth]{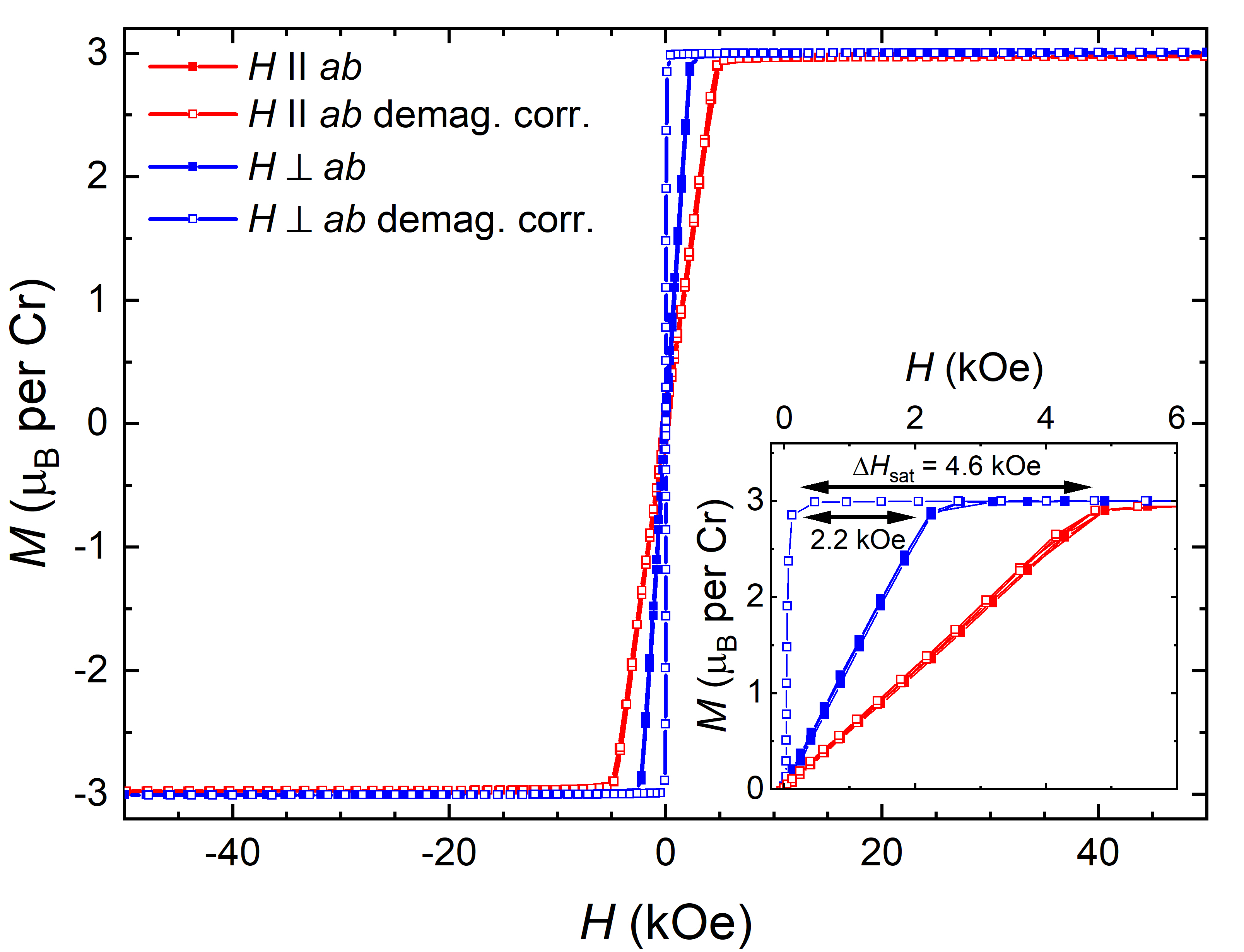}
\caption{Magnetization as function of field at 1.8\,K for both crystallographic orientations (open
symbols) and without (filled symbols) demagnetization field correction due to plate-like sample
shape (for details see text).} \label{fig:MH_1P8K}
\end{figure}

Fig.~\ref{fig:MH_1P8K} shows the isothermal magnetization of \CGT\ at 1.8\,K for $H \parallel ab$
and $H \perp ab$. The hysteresis of the magnetization as function of field is negligible, showing
the behavior expected for a soft ferromagnet. From the high-field region, a saturation
magnetization of $M_S \approx 3\,\mu_{B}$/Cr is obtained for both orientations. Thus, Cr$^{3+}$ with
S\,=\,3/2 leads to an isotropic Land\'e factor of g\,$\approx$\,2, which is in excellent agreement
with recent results from FMR studies on this compound\citep{JZeisner2019}. The saturation field is found as the \textit{x}-component of the intercept of  two linear fits, one being a fit to the saturated regime at high fields and one being a fit of the unsaturated linear regime at low fields. While the saturation magnetization is isotropic, the saturation field is anisotropic and changes from $H_{sat} = 4.8$\,kOe for $H
\parallel ab$ to $H_{sat} = 2.3$\,kOe for $H \perp ab$.\\
This anisotropic behavior in the isothermal magnetization is related to two different contributions: the intrinsic magnetic anisotropy of the
material (magnetocrystalline anistropy) and the shape anisotropy of the measured sample. As \CGT\
grows as thin platelet crystals, the shape anisotropy must be explicitly taken into account. To evaluate the demagnetization factors the sample's dimensions were measured along its edges from which an equivalent cuboid was constructed. The demagnetization factors of $N_x$\,=\,$N_y$\,=\,0.06 and $N_z$\,=\,0.88 were then calculated based on the equivalent-ellipsoid method\citep{JOsborn1945,DCronemeyer1991}.\\
As seen in Fig.~\ref{fig:MH_1P8K}, this correction strongly reduces the saturation field to 0.1\,kOe for the orientation $H \perp ab$, while only a negligible shift to 4.7\,kOe is obtained for $H \parallel ab$. The remaining anisotropy is purely originating from the magnetocrystalline anisotropy, showing that the magnetocrystalline easy axis is perpendicular to the crystallogrpahic \textit{ab}-planes (or in turn parallel to the \textit{c}-direction).\\

Using the Stoner-Wolfarth model\citep{EStoner1948} a value for the magnetocrystalline anisotropy
constant $K_U$ can be estimated from the saturation regime in the isothermal magnetization curve. Within
this model the magnetocrystalline anisotropy in the single domain state is related to the
saturation field $H_{sat}$ and the saturation moment $M_S$ with $\mu_0$ being the vacuum
permeability:
\begin{equation}
\frac{2K_U}{M_S} = \mu_0 H_{sat}
\label{eq:KU_Stoner_Wolfarth}
\end{equation}
For $H \parallel ab$, where the anisotropy becomes maximal, this yields $K_U = 47\pm1$\,kJ/m$^3$ at 1.8\,K. This value of $K_U$ is in good agreement with $K_U$ obtained previously by FMR on \CGT\citep{JZeisner2019}.\\

In general, it can be expected that the anisotropic anomaly observed in temperature dependent magnetization also manifests in the field dependence for $H \parallel ab$ (via a change of slope). Such a behavior was not resolved in our
data at 1.8\,K. This can be explained by the field dependence of \textit{T*}, which is investigated
in detail in the following subsection.

\subsection{Influence of external fields on the ground state}

\begin{figure*}[!htbp]
\centering
\includegraphics[width=\linewidth]{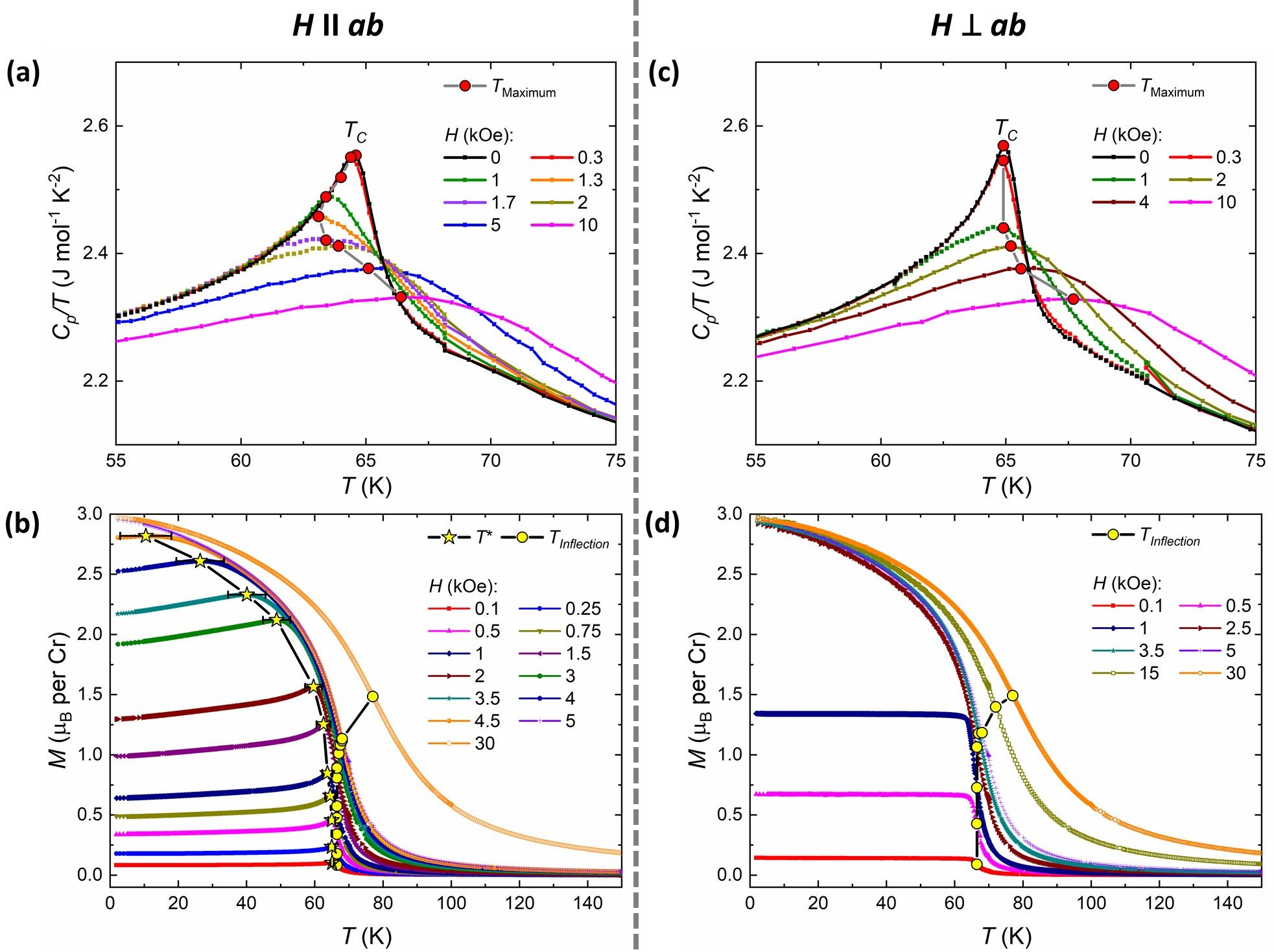}
\caption{Left: (a) $C_p/T$ and (b) Magnetization $M$ as a function of temperature under different
magnetic fields applied along the easy magnetization axis $c$. Right: (c) $C_p/T$ and (d) Magnetization $M$ as a function of temperature under different magnetic fields applied in the hard magnetization plane $ab$. The maxima in $C_p/T$ are marked with red dots in (a) and (c). The inflection points in \textit{M(T)} are marked with yellow dots in (b) and (d), while yellow stars in (b) indicate the maxima observed in \textit{M(T)} with $H \parallel ab$ corresponding to \textit{T*}.} \label{fig:Field_dep}
\end{figure*}

For $H \perp ab$ (Fig.~\ref{fig:Field_dep}(c) and Fig.~\ref{fig:Field_dep}(d)) the usual field dependence of ferromagnetic materials
is observed. In our specific heat studies the $\Lambda$-shape peak at $T_C$ evolves into a broad
maximum indicating that the magnetic transition becomes a crossover and this crossover is slightly
shifted to higher temperature under magnetic fields. This is in agreement with the change seen in
the temperature dependent magnetization curve.\\

Overall, a different behavior is seen for $H \parallel ab$ (Fig.~\ref{fig:Field_dep}(a) and Fig.~\ref{fig:Field_dep}(b)). While in the
magnetization curves the ferromagnetic phase transition at $T_C$ behaves in a similar way, for the
lowest measured field of 0.1\,kOe the downturn of the magnetization towards lower temperatures sets
in just below the Curie temperature. This is indicated by a maximum in the magnetization curve at
\textit{T*} (Fig.~\ref{fig:Field_dep}(b)). By increasing the external field, \textit{T*} shifts towards
lower temperatures. Additionally, upon increasing the external field, not just \textit{T*} shifts
towards lower temperatures but also the maximum itself gets broadened and the downturn itself gets
less pronounced. Finally at 5\,kOe, which is close to the saturation field for the hard
magnetization plane \textit{ab}, no downturn is obtained anymore. Furthermore, comparing the
temperature dependent magnetization for $H \parallel ab$ and $H \perp
ab$ at 5\,kOe or higher fields, anisotropic magnetization is observed.\\
In comparison to the magnetization data, the specific heat only shows one clear phase transition
for $H \parallel ab$, together with a change of the shift of the $\Lambda$-shaped peak position
around 1.7 kOe (\ref{fig:Field_dep}(a)). By increasing the external field from zero up to 1.3\,kOe the
position of the maximum shifts towards lower temperatures. By increasing the external field
further, the position of the maximum starts to shift towards higher temperatures until an isotropic
behavior is observed for fields of 5\,kOe and higher. Furthermore, the  progressive broadening of the maximum of $C_p/T$ indicates an evolution of the nature of the transition from a second-order phase transition to a crossover.\\
Considering the strength of the downturn, seen in the temperature dependent magnetization for
$H \parallel ab$, an observable entropy change is expected to go along with its onset. Therefore a
corresponding anomaly in $C_p/T(T)$ is expected. In the field range of 0\,kOe and 1.3\,kOe only one
distinct signal is found in $C_p/T(T)$. However, in this field range \textit{T*} and $T_C$ are
close to each other (less than 3\,K difference) and the $\Lambda$-shaped signal in the specific
heat has a significant broadness. Therefore, it is not possible to state if only one anomaly is
observed or if the signal contains actually two anomalies in this field range. However, as the
signal in specific heat shifts towards lower temperatures, a dominant influence of the transition
at \textit{T*} in this field regime can be expected.\\

While the crossover resulting from the PM-FM transition shifts towards higher temperatures as seen
for $H \perp ab$ for both physical properties $M$ and $C_p$, for $H \parallel ab$ and low fields
the dominating \textit{T*} shifts towards lower temperatures and is illustrated via an anomaly in
$C_p/T(T)$. For fields in the range of 1.7\,kOe to 5\,kOe, however, the specific heat measurements
clearly show the absence of entropy changes at \textit{T*} but seems again to be sensitive to
changes at $T_C$. This indicates that \textit{T*} is a transition between two states with comparable magnetic entropy. At fields above 5\,kOe the specific heat behavior is isotropic for fields parallel and
perpendicular to \textit{ab}. This is in good agreement with the fields found for isotropic
behavior in the temperature dependent magnetization.

\subsection{Low-Field Magnetic Phase Diagrams}

\begin{figure*}[p]
\centering
\includegraphics[width=0.7\linewidth]{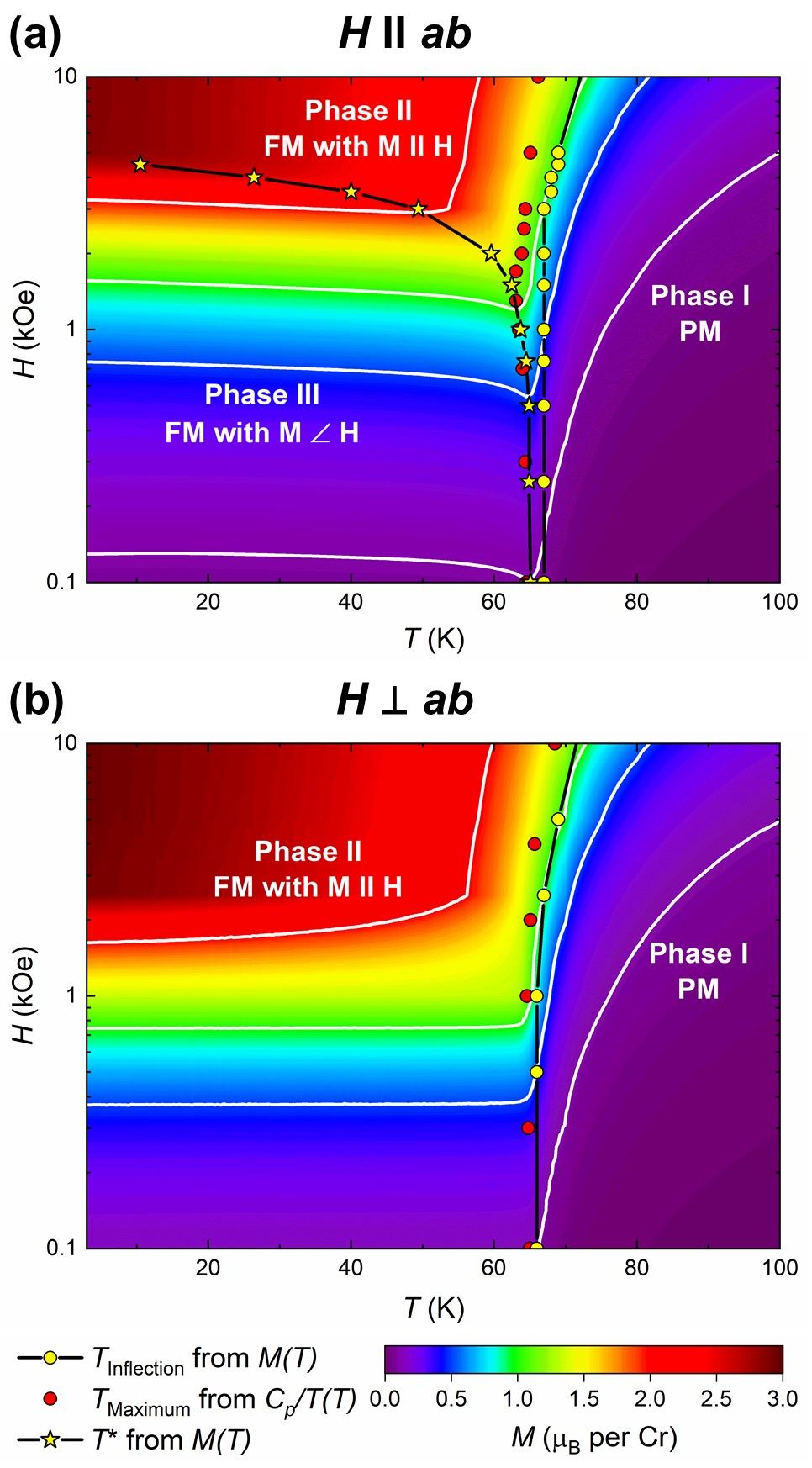}
\caption{Low-field magnetic phase diagram of \CGT\ for (a) $H \parallel ab$ and (b) $H \perp ab$,
where Phase I is the paramagnetic state; Phase II is the ferromagnetic state with $M \parallel H$;
Phase III only for $H \parallel ab$ is the ferromagnetic state with $M \protect\angle H$ due to the
interplay between $K_{U,eff}$, $H$ and $T$ as schematically shown in Fig.
\ref{fig:Spin_Orientation_scheme}. For both phase diagrams iso-magnetization lines at 0.1\,$\mu_B$, 0.5\,$\mu_B$, 1\,$\mu_B$ and 2\,$\mu_B$ are shown in white. The legend and the color scale at the bottom are applicable to both phase diagrams. Note that the magnetization shown in the phase diagrams is only the magnetization component parallel to \textit{H}.}
\label{fig:magnetic_phasediagrams}
\end{figure*}

For a better comparison between the peak position in specific heat and the significant temperatures
from magnetization, the low-field magnetic phase diagrams for fields along the easy axis and the
hard plane were constructed from our data. For fields along the easy axis (\ref{fig:magnetic_phasediagrams}(b)),
two phases are seen, i.e., a disordered paramagnetic phase (Phase I) at high temperatures and a
ferromagnetic ordered state with $M \parallel H$ (Phase II) at lower temperatures. The transition
temperatures from specific heat (peak position) and from magnetization (inflection point) are in
good agreement within the range of the measurement uncertainties. In zero field the magnetization
direction is supposed to be along the easy axis in the ferromagnetic state. Applying external
fields parallel to the magnetic easy axis stabilizes this state for example against thermally
activated magnetic fluctuations. Therefore, the observed behavior of Phase II as function of field
and temperature is well expected.\\

However, for $H \parallel ab$ an additional Phase\,III is observed, as shown in Fig.~\ref{fig:magnetic_phasediagrams}(a). While for $H \perp ab$ the iso-magnetization lines are parallel to the T-axis until they deviate towards higher fields very
close to $T_C$ , for $H \parallel ab$, these lines first show a trend towards lower fields before
they finally deviate towards high fields at elevated temperatures. These kinks are the fingerprints
of the maximum seen in the temperature dependent magnetization and are well followed by
\textit{T*}. This allows to not just define \textit{T*(H)} but also \textit{H*(T)} in this low temperature/low field regime. Whereas \textit{T*(H)} corresponds to the signature of Phase\,III in temperature dependent magnetization, \textit{H*(T)} corresponds to the same signature in field dependent magnetization. Using the magnetic phase diagram for $H \parallel ab$ to estimate \textit{H*}(1.8 K) explains why no anomaly could be resolved in the corresponding isothermal magnetization in Fig.~\ref{fig:MH_1P8K}, as mentioned before. \textit{H*}(1.8 K) is estimated to be in the range of 4.5\,--\,4.7\,kOe which is close to the saturation magnetization at this temperature. Consequently the slope of the \textit{M(H)} curve significantly changes in this field range and a separate anomaly corresponding to the signature of Phase III is not resolved.\\
Besides the low temperature/low field regime (Phase III) which is separated from the
rest of the phase diagram by \textit{T*}, both phase diagrams resemble each other. This is best
seen by comparing the course of the iso-magnetization lines outside of Phase III. Consequently, the
magnetization in Phase I and II is considered as isotropic and the direction of the magnetization
is parallel to the field for $T\textrm{*} < T < T_C$ as seen for $H \perp ab$ (Phase\,II).

\begin{figure}[!htbp]
\centering
\includegraphics[width=\linewidth]{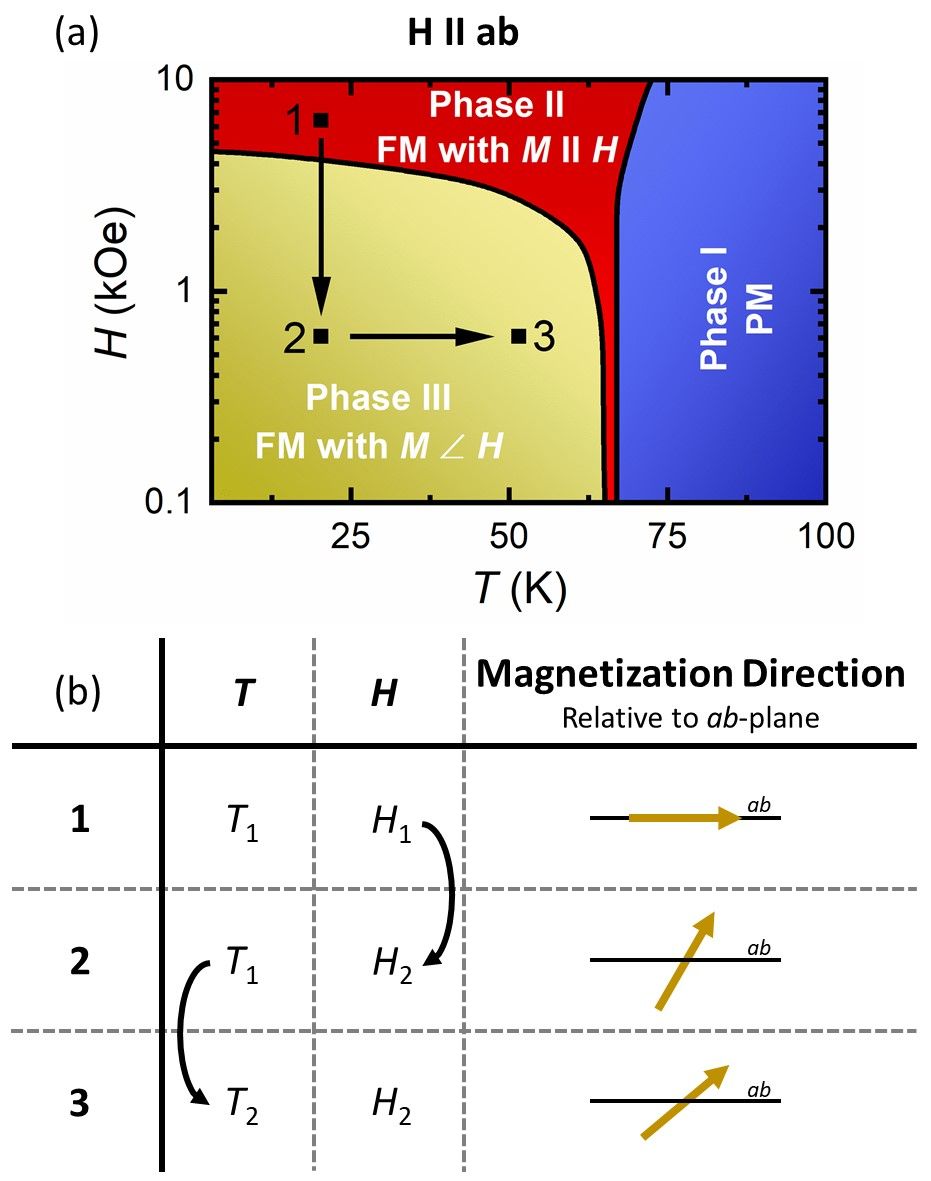}
\caption{(a) Schematic representation of the magnetic phase diagram of \CGT\ for $H \parallel ab$ with the three magnetic phases as shown in Fig.~\ref{fig:magnetic_phasediagrams}(a) with three points indicated. These points are arbitrarily chosen, however, fulfill the following conditions: $T_1 < T_2 < T\textrm{*}(\textrm{100 Oe)}$ and $H_1 > H\textrm{*}(T_1) > H_2$. (b) shows the parameters \textit{T} and \textit{H} for every point together with the expected direction of the magnetization with respect to the \textit{ab}-plane. The black arrows in (a) correspond to the arrows in (b) and indicate which of the parameters \textit{T} and \textit{H} is changed. Please note, that the arrow for the magnetization direction is supposed to only show the direction of the magnetization vector and not its absolute value.}
\label{fig:Spin_Orientation_scheme}
\end{figure}

Concluding from this behavior, the most likely scenario for the origin of the downturn in the
magnetization curve for $H \parallel ab$ is a continuous reorientation of the magnetization
direction as result of an interplay between the magnetocrystalline anisotropy, field and
temperature, as schematically shown in Fig.~\ref{fig:Spin_Orientation_scheme}. The
magnetocrystalline anisotropy favors a magnetization direction perpendicular to \textit{ab}, while
for $H \parallel ab$ the field wants to align the magnetization direction parallel to the field.
Assuming an external field $H_1$ that is higher than \textit{H*} at a given temperature $T_1$, the magnetization vector is aligned along the field direction (Point 1 in Fig.~\ref{fig:Spin_Orientation_scheme}).
However, by reducing the external field to $H_2$ below \textit{H*} at the same temperature, a tilting of the magnetization vector away from the
field direction will be achieved, i.e., a tilting towards the easy axis $c$ in this case (Point 2 in Fig.~\ref{fig:Spin_Orientation_scheme}). This is due to the reduction of external field leading to similar energy scales of the magnetic field and the magnetocrystalline anisotropy. The tilting in turn leads to a reduction of the magnetization component parallel to the field ($\perp ab$ in this case).\\

In order to follow and describe this effect as function of temperature, a temperature dependent
magnetic anisotropy has to be taken into account. The magnetocrystalline anisotropy is caused by
the underlying crystallographic lattice which is connected to the electronic spins via the
spin-orbit coupling. As such, the magnetocrystalline anisotropy constant $K_U$ is considered as a
material constant which itself is independent of temperature and field.\\
However, if the underlying lattice deforms anisotropically as function of temperature, also $K_U$ changes as a result. In
\CGT\ such an anisotropic temperature dependence of the lattice was observed by Carteaux \textit{et
al.}\citep{VCarteaux1995}. Down to 100 K the lattice parameters \textit{a} and \textit{c} shrink monotonously. However, around 100\,K the \textit{a}-axis starts to increase towards lower temperatures while the \textit{c}-axis shrinks further. The increase of the \textit{a}-parameter leads to a value of 6.820\,\AA\ at 5\,K which is larger than 6.812\,\AA\ at 270\,K. The temperature-onset of the increase of the \textit{a}-axis with approximately 100\,K agrees well with the temperatures which showed first low dimensional magnetic contributions to the linewidth in ESR experiments\citep{JZeisner2019} and with $\Theta_{CW}$ obtained in Fig.~\ref{fig:basic_magnetic_characterization}(c). Consequently, a connection between the onset of ferromagnetic interactions and the anisotropic behavior of the lattice parameters in form of magnetostriction may be expected. Most probably, however, this behavior is not sufficiently strong to explain the observed anomaly in magnetization.

\begin{figure}[!htbp]
\centering
\includegraphics[width=0.9\linewidth]{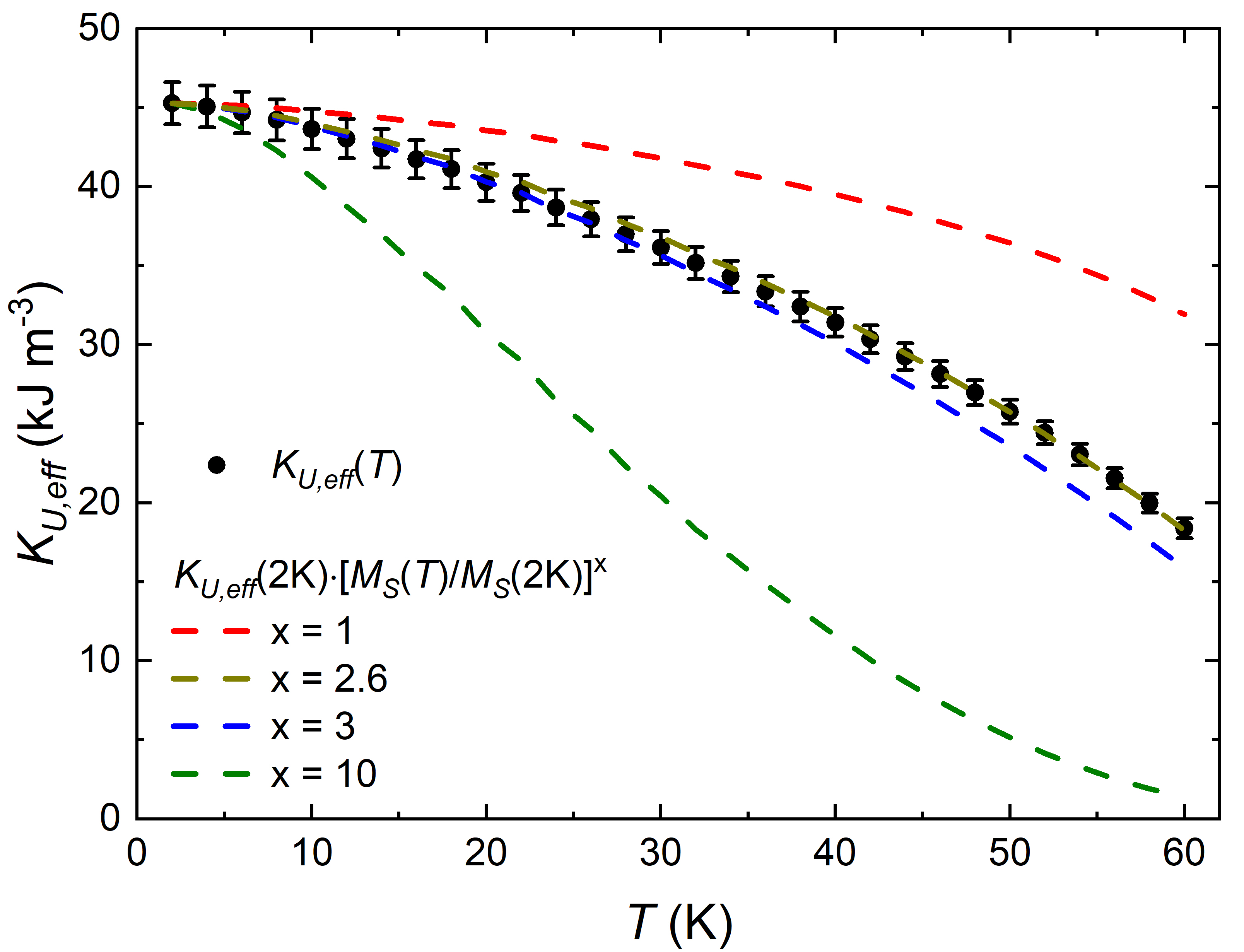}
\caption{Temperature evolution of the effective magnetic anisotropy constant $K_{U,eff}$ and the
expected scaling of $K_{U,eff}$ according to the power law behavior described in Eq.~\ref{eq:Anisotropy_Powerlaw} using the exponents 1, 2.6, 3 and 10.} \label{fig:KUeff_T}
\end{figure}

C. Zener \citep{CZener1954} described the effect of temperature fluctuations on the anisotropy of
the magnetization. According to his work, temperature leads to independent random fluctuations of
local magnetization directions. In turn, this leads to an effective reduction of both macroscopic
magnetization and anisotropy in the system. However, on a local scale the magnetization and
magnetic anisotropy are temperature independent. To differentiate between the local temperature
independent and the global temperature dependent magnetic anisotropy, K$_{U,eff}$ is introduced as
an effective anisotropy constant which includes the effect of thermal fluctuations on a macroscopic
scale and its interplay with the temperature independent $K_U$. Based on Eq.
\ref{eq:KU_Stoner_Wolfarth}, the temperature evolution of K$_{U,eff}$ was extracted from the
magnetic phase diagram with fields parallel to the \textit{ab}-plane and is represented in Fig.
\ref{fig:KUeff_T}. Details of how $K_{U,eff}$ was obtained are given in the Appendix \ref{supp}.\\
As both the macroscopic magnetization and anisotropy are affected by thermal fluctuations, a
proportionality between their evolution as function of temperature can be expected. According to
the theory by H.B.\,Callen and E.\,Callen\citep{HCallen1966}, this proportionality can be expressed
by a power law behavior of
\begin{equation}
\frac{K_{U,eff}(T)}{K_{U}} = \biggl[\frac{M_S(T)}{M_S}\biggl]^{\frac{l(l+1)}{2}}.
\label{eq:Anisotropy_Powerlaw}
\end{equation}
Hereinafter, the approximations $K_U \approx K_{U,eff}(2\,\textrm{K})$ and $M_S \approx
M_S(2\,\textrm{K})$ are used. In the case of uniaxial anisotropy $l = 2$ and an exponent of 3 are
expected, while for cubic anisotropy $l = 4$ and an exponent of 10 are found.\\
Fig.~\ref{fig:KUeff_T} shows the expected evolution of K$_{U,eff} (T)$ given by the power law dependence
of the saturation magnetization in Eq.~\ref{eq:Anisotropy_Powerlaw} for exponents 1, 2.6, 3 and 10.
The observed temperature dependence of K$_{U,eff}$ at low temperatures shows a good agreement with
the Callen-Callen power law with an exponent of 3, which is expected for purely uniaxial
anisotropy. However, at higher temperatures the exponent deviates from 3 towards 2.6. This
deviation can most probably be attributed to the change of $K_U$ itself due to lattice deformations
as result of magnetostriction as introduced before. For the exponents 1 and 10 the power law behavior does not follow K$_{U,eff} (T)$ and therefore direct scaling of the saturation magnetization with K$_{U,eff}$ as well as cubic anisotropy can be ruled out.\\
This confirms that the magnetic anisotropy in \CGT\ is uniaxial as expected given the non-cubic crystal structure, and the good agreement of simulations and experimental values of the angular dependence of the resonance field in FMR using an uniaxial model
in our previous work\citep{JZeisner2019}. Therefore, also the observed reduction of the magnetic
anisotropy as function of temperature seems to be reliable.\\
It should be noted that S.\,Khan\,\textit{et\,al.} also reported a temperature dependent $K_{U,eff}$ for \CGT\ which,
however, scales with an exponent 4.71\citep{SKhan2019}. They proposed that this deviation from the
expected exponent of 3 is due to the role of spin-orbit coupling from Te atoms, which is not
observed in our study. Furthermore, our analysis is very similar to N.\,Richter\,\textit{et\,al.} on CrI$_3$\citep{NRichter2018}, who also do not see a significant role of spin-orbit coupling on the temperature dependent
$K_{U,eff}$ values in their compound.\\

Assuming a tilted magnetization vector due to the previously discussed interplay between the
effective magnetic anisotropy and an external field perpendicular to the easy axis at $T_1$ (Point 2 in Fig.~\ref{fig:Spin_Orientation_scheme}), an increase in temperature to $T_2$ leads to a reduction of magnetic anisotropy. Therefore, the
alignment along the magnetic easy axis becomes less favorable upon warming, which leads to a
stronger tilting of the magnetization vector towards the \textit{ab}-plane and an increased
experimentally determined \textit{ab}-component $\parallel H$ in this case (Point 3 in Fig.~\ref{fig:Spin_Orientation_scheme}).\\
Thus, \textit{T*} is the temperature at which the magnetization component along the easy axis becomes finite upon
decreasing temperatures at a constant field in the \textit{ab}-plane. \textit{Vice versa}, \textit{H*} is the field in the \textit{ab}-plane below which the easy axis magnetization component becomes finite at a constant temperature. A similar scenario was
already proposed to explain a similar downturn of the transverse magnetization upon cooling below
the Curie temperature in other ferromagnets: the structurally related quasi two-dimensional
ferromagnets CrX$_3$ (X = Br, I)\citep{NRichter2018} and the heavy Fermion ferromagnet
URhGe\citep{FHardy2011}.

\begin{table}
\begin{tabular}{ccc}
\hline \hline
Compound & K$_U$ [kJ/m$^{3}$] & Reference\\
\hline
CrBr$_3$ & 86 ($\pm6$) & N.~Richter~\textit{et al.}\citep{NRichter2018}\\
CrI$_3$ & 301 ($\pm50$) & N.~Richter~\textit{et al.}\citep{NRichter2018}\\
\CGT & 47 ($\pm1$) & this work\\
\hline \hline
\end{tabular}
\caption{Comparison between $K_U$ for different (quasi-)2D honeycomb ferromagnets. Please note that
for CrBr$_3$ and CrI$_3$ $K_U$ was extracted from isothermal magnetization data at $T = 5$\,K while
for \CGT\ data at $T = 1.8$\,K was used.} \label{tab:Comparison_KU_CGT_CrX3}
\end{table}

For CrX$_3$ (X = Br, I) a similar analysis of $K_{U,eff} (T)$ was performed\citep{NRichter2018}. While the
magnetocrystalline anisotropy constants of the chromium halides are larger than the one found for
\CGT\ (shown in Tab.~\ref{tab:Comparison_KU_CGT_CrX3}), their temperature dependence is also well
described by exponents according to an uniaxial anisotropy. In the case of URhGe, the tilting of
the magnetic moment in between the field direction and the easy magnetization axis was directly
observed by neutron diffraction\citep{FLevy2005} and NMR\citep{HKotegawa2015}. For URhGe a Ginzburg
Landau description of the anisotropic ferromagnet proposed by V. Mineev \citep{VMineev2011}
reproduced the downturn of the magnetization and could possibly also be a promising model for a
simple description of the low-field magnetic properties of \CGT.

\section{Summary}
\label{Summary}

In summary, detailed magnetic and thermodynamic measurements were performed on high-quality \CGT\
single crystals. Analysis of the low field data shows an interesting interplay of $K_U$, applied
magnetic field and temperature. \CGT\ is a soft ferromagnet with a Curie temperature $T_C=65$\,K.
An effective moment $\mu_{eff}\approx4\mu_B$/Cr and an isotropic saturation moment $M_S =
3$\,$\mu_B$/Cr were found, both being in good agreement with the values expected for Cr$^{3+}$.
Furthermore, the isotropic saturation magnetization hints towards an isotropic Land\'{e}-factor $g
\approx 2$. The difference between $\Theta_{CW}=95$\,K and $T_C$ as well as the shape of the
temperature dependent specific heat indicate low-dimensional magnetic fluctuations well above the
magnetic ordering temperature. The easy-axis nature of the magnetic properties perpendicular to the
structural layers in the \textit{ab}-plane is confirmed and a magnetocrystalline anisotropy
constant $K_U = 47\pm1$\,kJ/m$^3$ is obtained using the Stoner-Wolfarth model.\\

The field and temperature dependence of the magnetization was studied in detail for fields parallel
and perpendicular to the hard magnetic plane \textit{ab} up to fields of 30\,kOe. Corresponding
magnetic phase diagrams were constructed. The field and temperature dependence for fields along the
easy axis $\parallel c$ show the typical behavior of a ferromagnet. However, for fields applied in
the hard plane \textit{ab} below a temperature $T^* < T_C$ a downturn towards lower temperatures is
found in magnetization curves below the saturation field $H_{sat,ab}\approx 5\,kOe$. The origin of
this anisotropic anomaly is discussed in terms of an interplay between the effective magnetic
anisotropy $K_{U,eff}$, temperature and the applied magnetic field. In this scenario, the
magnetization direction continuously changes between a field-parallel configuration above
\textit{T*} to a tilted direction with a magnetization component perpendicular to $H$. Thus, the
temperature \textit{T*} can be understood as the temperature where the magnetization component
perpendicular to the \textit{ab}-plane changes from zero to finite.\\

To investigate the validity of the temperature dependence of the magnetic anisotropy, values for
$K_{U,eff}$ were extracted at different temperatures from the magnetic phase diagram for $H
\parallel ab$ and compared with a power law scaling of the temperature dependent saturation
magnetization according to H.B Callen and E. Callen\citep{HCallen1966}. The observed power law
behavior fits well for uniaxial anisotropy models with a small deviation at higher temperatures,
which can most probably be attributed to changes of $K_U$ itself due to temperature dependent
anisotropic lattice deformations.\\

A similar anisotropic anomaly was observed for CrX$_3$ (with X\,=\,Br,\,I) and also discussed in
terms of interplay between $K_{U,eff}$ and temperature\citep{NRichter2018}. All these compounds
share the same magnetic ion and easy axis $\parallel c$ ferromagnetic ordering together with a
similar 2D honeycomb lattice. Thus, the magnetocrystalline anisotropies in these systems are
similar, although the magnetocrystalline anisotropy constant $K_U$ shows significant differences in
its absolute value for the mentioned compounds. This hints towards a universality of this interplay
in quasi two-dimensional ferromagnetic materials. Furthermore, the observed anomaly in the temperature dependence of the magnetization can be considered as a fingerprint of this interplay.\\

Besides all similarities, these compounds also show influential differences in the nature of their
magnetism, for example in the type of magnetic coupling. The TMTCs (\CGT\citep{YLiu2017} and
Cr$_2$Si$_2$Te$_6$\citep{BLiu2016}) exhibit 2D Ising like behavior while CrX$_3$
(X\,=\,Br\citep{JHo1969},\,I\citep{YLiu2018}) display a more 3D Ising-like coupling, according to
investigations of the critical behavior of these compounds due to interlayer interactions present
at least in the bulk state. Taken the 2D nature and the high Curie temperature of \CGT, this
compound could be a highly promising low-dimensional ferromagnet to gain further insight into
low-dimensional ferromagnetism in general and for the use in ferromagnetic heterostructures.

\section{Acknowledgements}
S.S. acknowledges financial support from GRK-1621 graduate academy of the
DFG. G.B. acknowledges financial support from the European Union's Horizon 2020
research and innovation program under the Marie~Sk\l{}odowska-Curie grant agreement No.~796048. S.A. acknowledges financial support from Deutsche Forschungsgemeinschaft (DFG) via Grant No.~DFG~AS~523\textbackslash4-1. A.U.B.W. and B.B. acknowledge financial support from the DFG through SFB~1143~(project-id 247310070). 

\bibliography{Thermodynamic_Cr2Ge2Te6_Literature}

\appendix
\section{Supplementary}
\label{supp}

\subsection{Extraction of $K_{U,eff}$ as function of temperature}

\begin{figure}[!htbp]
\centering
\includegraphics[width=0.9\linewidth]{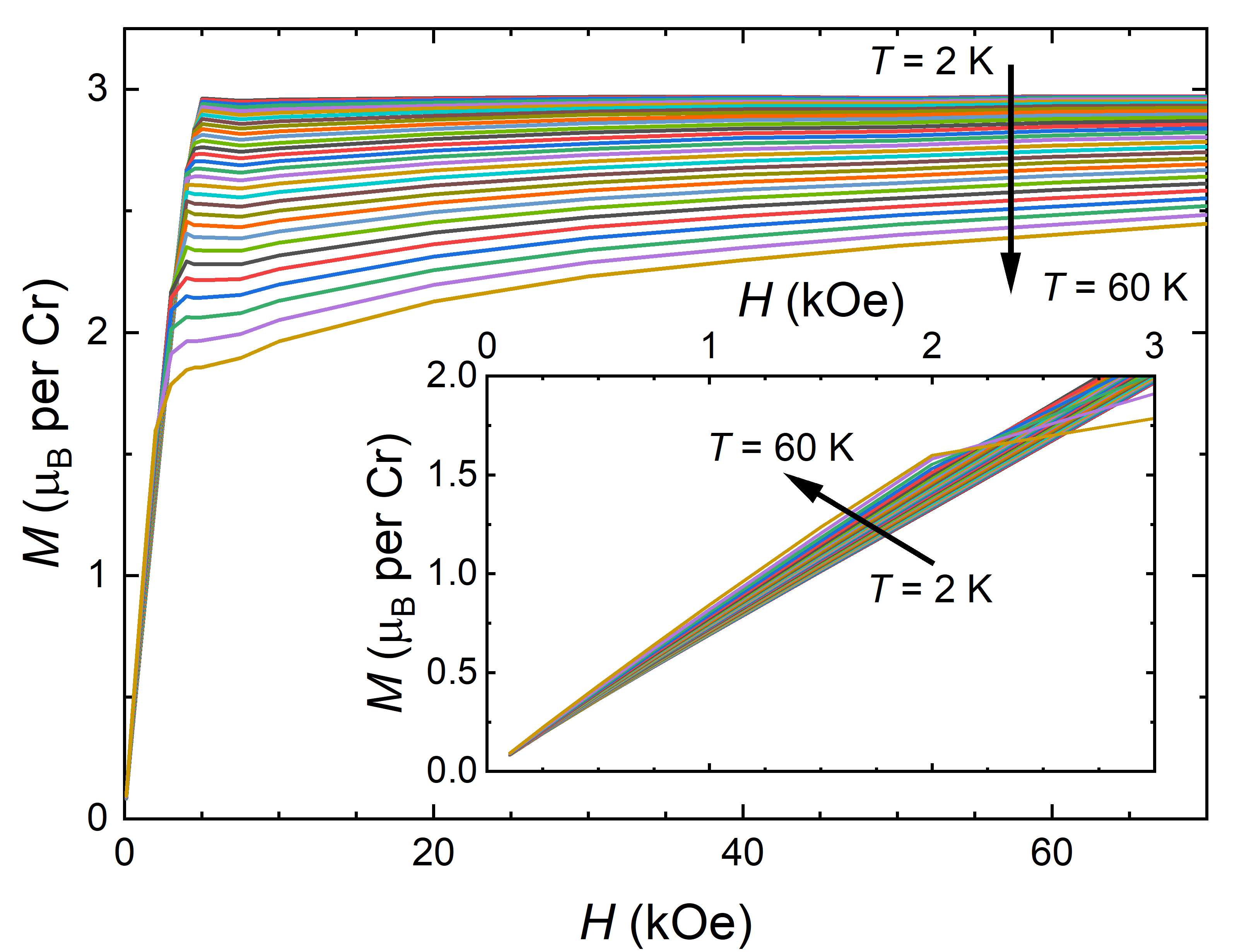}
\caption{Isothermal magnetization for $H \parallel ab$ in a range from 2\,K to 60\,K in 2\,K steps
extracted from the corresponding magnetic phase diagram. The inset shows the evolution of the linear behavior at low fields as function of temperature in more detail.} \label{fig:MvsH_from_phasediagram}
\end{figure}

To obtain the temperature dependence of $K_{U,eff}$, isothermal magnetization curves were extracted
from the magnetic phase diagram for $H \parallel ab$ in the range of 2\,K to 60\,K (shown in Fig.
\ref{fig:MvsH_from_phasediagram}). The saturation magnetization and the saturation field at each
temperature were obtained from the intersection of two linear regressions of the low-field (0\,kOe
to 2\,kOe) and the high-field region (30\,kOe to 70\,kOe), respectively. From these values
$K_{U,eff}$ was obtained for each temperature based on Eq.~\ref{eq:KU_Stoner_Wolfarth}.

\end{document}